\documentstyle[sprocl]{article}
\input{psfig}
\def\be{\begin{equation}}
\def\ee{\end{equation}}
\def\bea{\begin{eqnarray}}
\def\eea{\end{eqnarray}}

\begin{document}
\title{FCNC SOLUTIONS TO THE SOLAR NEUTRINO PROBLEM} 
\author{P.I. KRASTEV and J.N. BAHCALL} 
\address{Institute for Advanced Study,
Olden Lane, Princeton NJ 08540, U.S.A.}

\maketitle\abstracts{We present the status of FCNC solutions to the
solar neutrino problem.  Our analysis shows that the FCNC solutions
with massless neutrinos provide a good fit to the latest solar
neutrino data. Predictions for future experiments are formulated which
will permit further tests of these solutions.}

\section{Introduction}
In his seminal paper~\cite{wolf} Lincoln Wolfenstein considered flavor
changing neutral current (FCNC) effects on neutrino oscillations. He
noticed that if the neutrino current in the effective four-fermion
interaction Lagrangian has a flavor non-symmetric part it can
significantly affect neutrino oscillations by changing the neutrino
indices of refraction in matter. Furthermore, if the neutrinos are
massless, flavor diagonal neutral currents (FDNC), the couplings of
which are not flavor symmetric, are also needed for strong transitions
to take place.

In supersymmetric theories without R-parity, both FCNCs and FDNCs
occur at tree level. The FCNCs in these theories can enhance neutrino
oscillations in matter even for vanishingly small mixing angles in
vacuum~\cite{roulet}. In the minimal standard supersymmetric model
(MSSM) without R-parity the possibility exists~\cite{GMP} of solving
the solar neutrino problems (SNP) even with massless neutrinos by the
interplay of both FCNCs and FDNCs. The first definitive confrontation
with data, as well as the thorough investigation of all possibilities
done in ref.~\cite{BPHW} showed that this is indeed so.

Since 1991, when the FCNC solution was first considered, new solar
neutrino data has become available. In particular, reliable new data
from both gallium detectors, GALLEX and SAGE, now exists. The solar
models have been also significantly improved. The latest solar
models~\cite{BP95} include diffusion of helium and heavy elements,
which is important for the excellent agreement with
helioseismology~\cite{helio}.  These developments justify a new data
analysis in order to determine the status of the FCNC solution to the
SNP.

\section{FCNC mechanism}
\label{FCNC:mech}

In the MSSM without R-parity $\nu_e$ can scatter off $d$-quark via
exchange of a $b$--squark and either remain $\nu_e$ or emerge as a
$\nu_\tau$.  This tree level process leads to neutrino mixing in
matter even if the neutrinos are massless or the vacuum mixing is
zero.  In order for a resonant conversion between the neutrino flavors
to take place both FCNC and flavor diagonal but flavor non-symmetric
contributions to the neutrino indices of refraction need to be
present.  The relevant transitions can only take place between the
first and third generations because the data from rare decays, atomic
parity violation, deep-inelastic scattering etc. impose strong
constraints on the values of the flavor-changing Yukawa couplings in
the superpotential required for the mixing between the first two
generations in matter. In general there might be two types of
solutions depending on the type of quark, $u$ or $d$, which
facilitates the flavor changing transition. In the case of the MSSM
without R-parity, the parameters entering the evolution equations are
functions of the Yukawa couplings, $\lambda_{ijk}$, in the
superpotential and the mass of the $b$-squark:

\begin{eqnarray}
\epsilon_d & = &
\frac{\lambda^{'}_{331}\lambda^{'}_{131}}{4\sqrt{2}G_{F}m^{2}_{\tilde
b}}, \\ 
\epsilon^{'}_d & = &\frac{|\lambda^{'}_{331}|^2 -
|\lambda^{'}_{131}|^2}{4\sqrt{2}G_Fm^2_{\tilde b} }. 
\end{eqnarray}

The evolution equation for neutrino oscillations in matter does not
depend on the energy of the neutrinos. The resonance condition is:

\begin{equation}
\epsilon^{'}_d = \frac{1}{1 + 2N_n/N_e},
\end{equation}
where $N_n$ and $N_e$ are respectively the neutron and electron number
densities.  Unlike the MSW effect, both neutrinos and anti-neutrinos
can undergo simultaneously resonant transitions. Since the transitions
do not depend on the energy of the neutrinos, but only on the ratio of
the neutron to electron number densities, this poses a problem for
solving the SNP. As shown in ref.~\cite{const} any energy independent
suppression mechanism is ruled out by the current solar neutrino data
(assuming current solar models do not underestimate the uncertainties
in the solar neutrino fluxes) at 99.96 \% C.L. The solution of this
problem was pointed out in~\cite{GMP}. Since the different solar
neutrinos ($pp$, $^7{\rm Be}$, $^8{\rm B}$ etc.) have different
production regions in the sun, by properly choosing the parameter
$\epsilon^{'}_d$ the fluxes of these neutrinos can be depleted with
different suppression factors depending on the position of the
resonance in the sun.

\section{Present Status of the FCNC Solutions}
\label{CLplots}
\begin{figure} 
\hbox{\psfig{figure=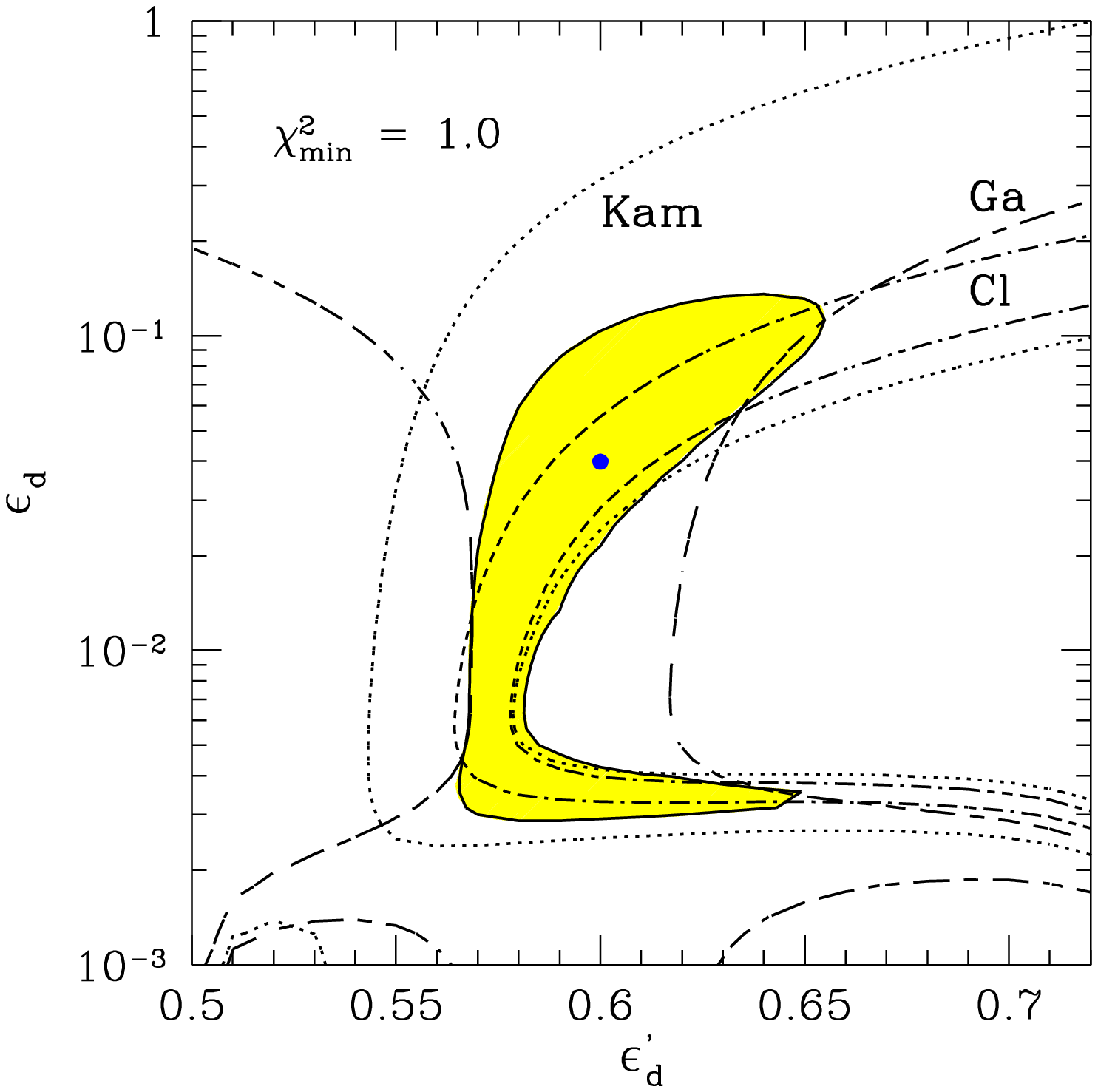,height=2.5in}
\hfill\psfig{figure=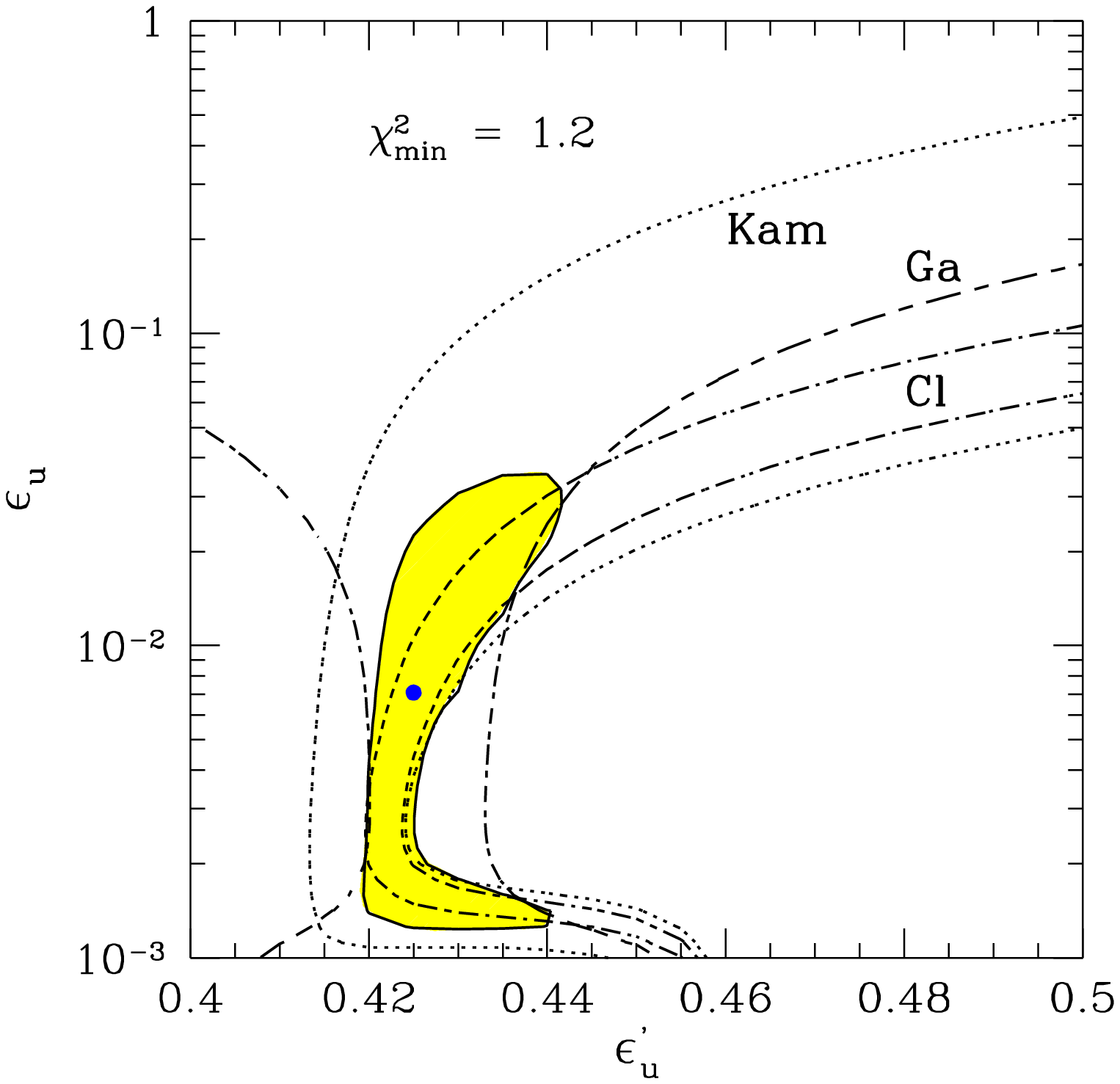,height=2.5in}} 
\caption{ Regions in parameter space allowed at 95 \% C.L. (shaded
area) from the latest (NEUTRINO '96) solar neutrino data. The regions
allowed by each solar neutrino experiment at 95 \% C.L. are also
shown.
\label{CL95}}
\end{figure}

In our analysis we use the latest solar neutrino data presented at the
NEUTRINO '96 conference. We compute the average survival probabilities
for $pp$, $^8{\rm B}$, $^7{\rm Be}$, $pep$ and CNO neutrinos to remain
electron neutrinos at the surface of the sun by averaging over their
respective production regions as given in the standard solar model
(SSM)~\cite{BP95}. Using the computed survival probabilities, the
solar neutrino fluxes from the SSM~\cite{BP95}, and published
detection cross-sections, we calculate the expected event rates in the
operating detectors. By a standard procedure, we calculate the
$\chi^2$ and determine the 95 \% C.L. allowed regions. The latter are
shown in Fig.\ref{CL95} for both the $d$- and $u$-quark solutions.

\section{Predictions for Future Experiments}
\label{future}

\begin{figure} 
\hbox{\psfig{figure=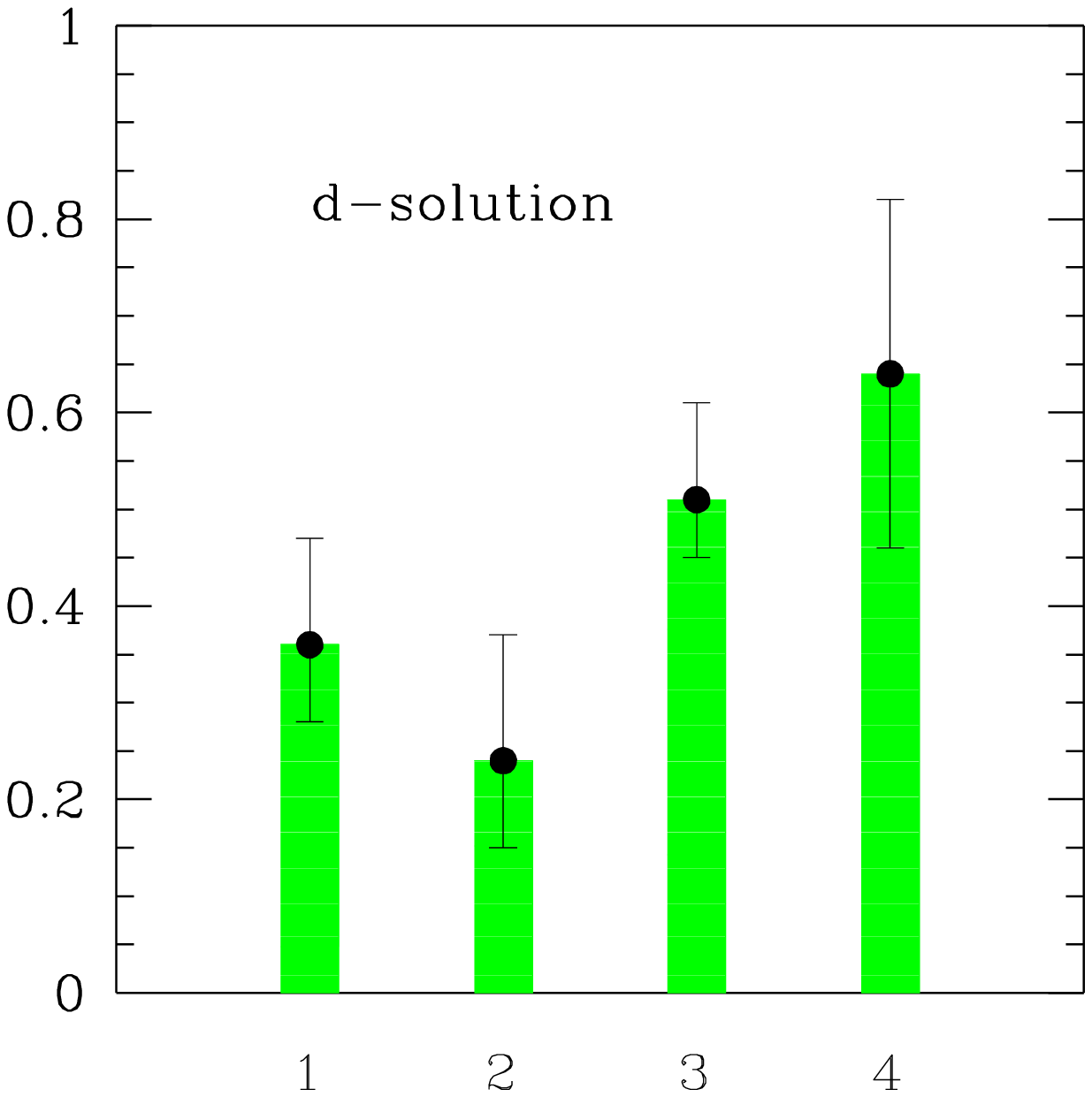,height=2.5in}
\hfill\psfig{figure=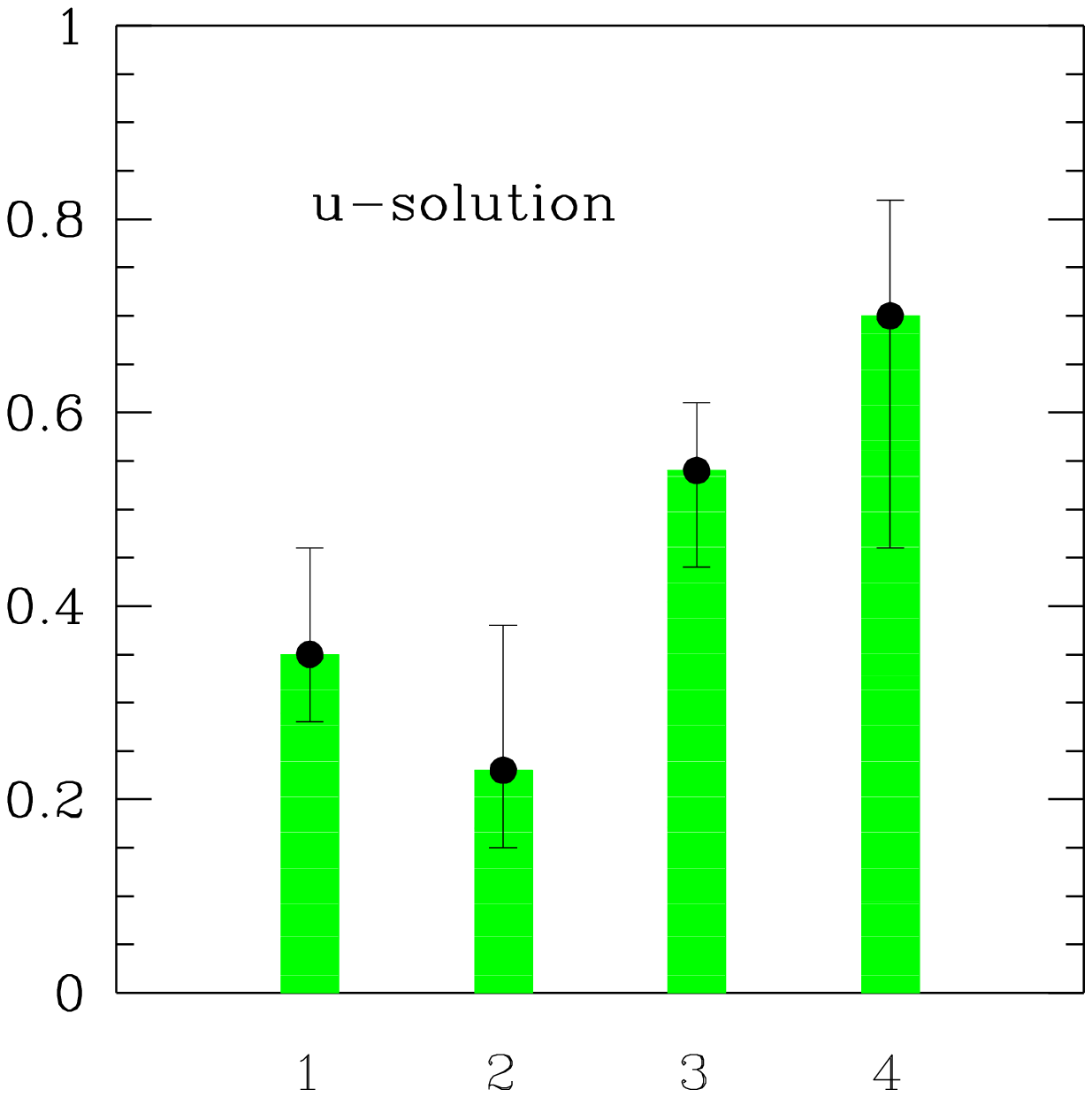,height=2.5in}} 
\caption{Predicted ratios of event rates after FCNC
($\nu_e\rightarrow\nu_\tau$) transitions to the corresponding expected
rates without transitions in: 1-SuperK, 2-SNO and ICARUS, 3-BOREXINO,
4-HELLAZ/HERON. The ranges in the event rates correspond to the
uncertainty in the parameters $\epsilon^{'}$ - $\epsilon$ which were
varied within the 95 \% C.L. allowed regions in Fig.\ref{CL95}.
\label{FD}}
\end{figure}

After obtaining the allowed at 95 \% C.L. regions in parameter space
we proceed with calculations of the event rates in future
detectors. The results of this calculation are summarized in
Fig.\ref{FD}. The ranges of the ratios of event rates reduced by FCNC
transitions in SuperKamiokande, SNO, BOREXINO, ICARUS and HELLAZ/HERON
to the corresponding event rates predicted with the neutrino fluxes
from the standard solar model by Bahcall and Pinsonneault~\cite{BP95}
are shown. In each case we have varied the parameters $\epsilon$ and
$\epsilon^{'}$ within the 95 \% C.L. region and for each point within
this region we have computed the predicted event rate in each of the
five detectors. An interesting prediction of the FCNC solutions is
that the signals in SNO and ICARUS should be suppressed by one and the
same factor even though the reactions in which solar neutrinos will be
detected in these two detectors are completely different.

Experiments that are being developed to test neutrino physics
solutions of the solar neutrino problem in a solar model independent
way will hopefully be able to distinguish also between different
neutrino physics solutions, e.g. between the MSW and FCNC
mechanisms. The combination of the three ``smoking gun'' effects,
namely distortion of the spectrum of recoil electrons in
neutrino-electron scattering experiments, the double ratio of CC/NC
events in SNO and the earth regeneration effect should look
differently in these two cases. The FCNC mechanism predicts no
distortion of the electron spectrum, whereas in the small mixing angle
(SMA) MSW solution this spectrum is expected to be distorted.  The
FCNC solutions also predict an energy-independent CC/NC ratio, similar
to the large mixing angle (LMA) MSW solution but unlike the SMA
solution where this ratio is expected to be energy dependent. Finally,
the night-day asymmetry due to the earth regeneration effect is
generally expected to be energy dependent for the MSW solution,
whereas the FCNC solutions predict a strictly energy independent
asymmetry.

\section{Conclusions}

The FCNC solutions with massless neutrinos arising from theories with
broken R-parity fit well the present solar neutrino data. However they
require unnatural choice of Yukawa couplings. Fine tuning is needed in
order to keep the neutrino masses and/or mixing angles negligibly
small~\cite{babu,yuval}.  Scenarios in which neutrinos have masses and
FCNCs and/or FDNCs contribute to the neutrino transitions in matter
are difficult to test experimentally since they usually introduce more
parameters and predict smaller deviations from either the vacuum
oscillations or the MSW mechanism. The latter currently provides the
best fit to the data among a variety of neutrino physics solutions of
the SNP without the necessity of any additional modifications of the
standard electroweak theory besides neutrino masses.

\section{Acknowledgments}

P. Krastev thanks K. Babu, Y. Grossman, E. Ma, S. Pakvasa and
S. Petcov  for discussions.

\end{document}